\documentclass[twocolumn,showpacs,preprintnumbers,amsmath,amssymb,prl]{revtex4}
\usepackage{graphicx,amssymb}
\usepackage{epsfig}

\begin{document}
\title{``Rectifying" reflection from a magnetic photonic crystal}
\author{Shiyang Liu$^1$, Wanli Lu$^1$, Zhifang Lin$^1$, and S. T. Chui$^2$}
\affiliation{$^1$ Surface Physics Laboratory, Department of Physics,Fudan University, Shanghai 200433, China\\
$^2$Bartol Research Institute and Department of Physics and
Astronomy, University of Delaware, Newark, DE 19716, USA}

\begin{abstract}
When an oscillating line source is placed in front of a special
mirror consisting of an array of flat uniformly spaced ferrite rods,
half of the image disappeared at some frequency. We believe that
this comes from the coupling to photonic states of the magnetic
surface plasmon band. These states exhibit giant circulations that
only go in one direction due to time reversal symmetry breaking.
Possible applications of this ``rectifying" reflection include a
robust one-way waveguide, a $90^\circ$ beam bender and a beam
splitter, which are shown to work even in the deep subwavelength
scale.
\end{abstract}


\maketitle

Anyone who has looked at himself in a mirror expects a faithful 
image from a localized source. As is
illustrated in Fig. 1(a -- c), we find a very surprising result that
when an oscillating line source is placed in front of a special
mirror consisting of a uniformly spaced flat collection of ferrite
rods (indicated by the black circles), the left half of the image
disappeared near some particular frequency.
We call this phenomenon ``Yin-Yang''
reflection (YYR), with ``Yin'' denoting the darkened region and
``Yang'' the brightened region.
The physics behind this is related
to the quantized Hall effect (QHE) and one-way magnetic surface
plasmon band states.

Electric currents that circulate around the edge of the sample are
believed to be responsible for the QHE.
Because of the broken time reversal symmetry (TRS) caused by the 
external magnetic field, this current only goes in one direction. 
There has recently been much interest in exploring if related
phenomena 
can
occur with photons. Among others, a skew scattering effect involving
electromagnetic (EM) waves was discussed by Rikken and coworkers
\cite{x}; possible edge-like one-way waveguide from magnetic
photonic crystal (MPC) bands with finite Chern numbers was discussed
by Haldane and Raghu \cite{Haldane} and Wang and coworkers
\cite{Joan}.

We have also been searching for similar phenomena involving EM waves
\cite{chui}. All QHE experiments deal with 
transmission. Because of the continuity of the electric and magnetic
fields at the boundary, photonic states with giant circulations can
be probed using reflection.  States with the largest one way circulation are
derived from the magnetic surface-plasmon (MSP) bands
\cite{msp1,lw}.

Plasmonic materials \cite{Ebsn,sp,konosky,Maier,abajo} are capturing
increasing interest due to their different promising applications.
Most attention until now has been focused on the electric surface
plasmons originating from the collective resonance of electronic
density wave and hosted by metallic building blocks. The symmetry of
Maxwell's equations with respect to the magnetic and electric
degrees of freedom enables a symmetric type of phenomenon 
in magnetic systems, which is known as a ``magnetic surface plasmon''
\cite{msp0}. When a periodic array of such material is assembled
together, the photonic states that hop from one MSP state to another
form a MSP {\bf band} \cite{msp1,lw}. In contrast to the electric
surface plasmon band, the MSP band states only go in one direction
(clockwise, for example), resulting from the breakdown of time
reversal invariance caused by the finite magnetic field. The YYR effect
comes from the coupling to states derived from these bands.

\begin{figure}
\includegraphics[width=7.9cm]{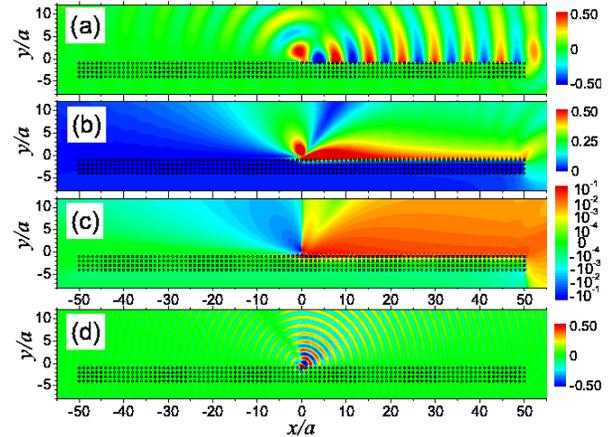}
\caption{\label{fig1} (color online).
The profile of the electric field with the sinusodial spatial dependence included (a), 
the amplitude of the electric field (b), and the $x$-component of
the Poynting vector (c) when a line source oscillating at a
frequency close to the magnetic surface plasmon resonance is placed
near the surface [(0,1.5a)] of a special mirror made of a slab
composed of a square lattice of YIG ferrite rods, indicated by the
black dots in the figure.
Figure (d) displays the electric field when the line
source oscillates at a frequency corresponding to the one-way edge mode
studied in \cite{Joan}. 
}
\end{figure}


The YYR effect has the potential to revolutionize microwave circuitry.
The diffraction limit requires that the minimum size of
waveguides be larger than half the wavelength of light. Surface
plasmons have attracted much attention as a way of circumventing the
diffraction limit. We show below that the YYR can be exploited to
build {\bf tunable} subwavelength waveguides, beam benders and beam splitters.
Furthermore, in these circuits, the EM wave exhibit a ``superflow"
behavior; the magnitude of the Poynting vector does not decrease in
the presence of different kinds of defects.
We next describe  in detail our results, which are obtained
using the rigorous multiple scattering method \cite{JOSA11,syliu}.

%

For definiteness, consider a MPC consisting of a square lattice of
parallel single crystal yttrium-iron-garnet (YIG) rods  along $z$ direction
in air separated by a distance $a$.
The radius of the YIG rod is $r=0.25a$. 
For single
crystal YIG, the saturation magnetization $M_0=1750$ gauss. The
damping coefficient $\alpha=3\times 10^{-4}$, and the permittivity
$\epsilon_s=15+3\times10^{-3}\,i$, corresponding to a gyromagnetic
linewidth of $0.3$ Oe and a dielectric loss tangent of $2\times
10^{-4}$, respectively \cite{Woh}. We focus on the transverse
magnetic (TM) wave with the electric field parallel to the rod axis.

Typical reflection behavior for $a= 8$ mm is shown in Figs. 1(a) and 1(b), where
the line source oscillating at a frequency of $5$ GHz is located
$1.5a$ away from a four-layer MPC slab. The applied external static
magnetic field is such that the total field is 900 Oe, 
corresponding to a MSP
resonance at $f_s=4.97$ GHz. The reflection is shaped such that the
reflected wave is dramatically different on the left hand side (LHS)
and the right hand side (RHS) of the source. On the LHS, the
scattered field largely attenuates the incoming field near the
interface. The MPC slab seems to repel the wave away from its
surface, resulting in a shadowy region, called the ``Yin" side. On
the RHS, the scattered field instead significantly boosts the incoming field
near
the interface. The MPC slab seems to attract 
the wave to its surface, giving rise to a brightened region, or a ``Yang'' side,
with an enhanced wave field.
We next discuss  the physics of the YYR.

Because of the broken TRS, the scattered field consists of unequal
amount of states with opposite angular momenta $n$ and $-n$,
resulting in a giant circulation. This unequal ratio is maintained
as the sign of the wave vector is changed from $(k_x,k_y)$ to
$(-k_x,k_y)$, so that the helicity of the circulation is unchanged.
As a result, reflection from the LHS and the RHS of the source will
be different. This difference is particularly dramatic only when the
frequency is near the MSP resonance, where the angular momenta
content is dominated by one sign, with the other sign almost
completely suppressed leading to a very large circulation
\cite{chui}. Consequently, only energy flow in one direction is
supported, while that in the other (opposite) direction is
substantially suppressed near the MPC surface. Coupling to these
states results in a brightened region on one side and a shadowy
region on the other side.
This is manifested in Fig.
1(c), where the $x$ component of the Poynting vector is exhibited.
The rightward energy flow is sustained and reinforced, while the
leftward energy flow is substantially suppressed, and eventually the
system ends up with a YYR phenomenon. With this type of reflection,
the EM wave is shaped to move mostly rightward near the interface.
Wang and coworkers \cite{Joan} recently demonstrated one-way
waveguide effects based on the edge state of the MPC
at a frequency regime away from the MSP. The asymmetry in
the reflected field is not as significant, as is shown
in Fig. 1(d).
Aside from the variation due to the Bragg reflections, there is only
a slight asymmetry in the field pattern near the MPC surface. This
demonstrates the importance of the MSP resonance in enhancing the
asymmetry of the reflection. We suspect that the MSP band states can
also enhance the effect of possible edge state waveguides.

Now we turn to explore the possible applications of the YYR
phenomenon. The first example is based on the remarkable asymmetry
in the Poynting vector as shown in Fig. 1(c). Because there are band
gaps both above and below the MSP resonance, the EM wave can be
confined between two MPC slabs, similar to the conventional photonic
crystal waveguide \cite{pcwg}. If the magnetization of the two MPC
slabs are in opposite directions, then the EM wave reflected forward
from one MPC slab will also be reflected forward from the other MPC
slab, leading to a different design of a one-way EM waveguide
\cite{Haldane,Joan,yu}, as is illustrated in Fig. 2(a). In Fig. 2, a
line source is placed at $(-5.5a,0)$ between two MPC slabs $2a$
apart. The parameters for the MPC slabs as well as the oscillating
frequency of the line source are the same as in Fig. 1(a -- c). It
can be seen that the wave propagates rightward.
Fig. 2(b)  shows the $x$ component $P_x$ of the Poynting vector
$\boldsymbol{P}$ in a logarithmic scale.
To further demonstrate the one-way transport characteristic, in Fig. 2(c) we display
the electric field along $y=0$ and the rightward transmitted power, $T_x 
=\int_{-3a}^{3a}P_x\text{d}y$,
as a function of $x$. 
In addition, the EM flux exhibits an extremely low decay rate,
the propagation loss is less than $0.002$ dB$/\text{mm}$, using
realistic material parameters for commercially available YIG ferrite
\cite{Woh}. To illustrate the physics, in the following, we shall
neglect the damping.


\begin{figure}
\includegraphics[width=7.9cm]{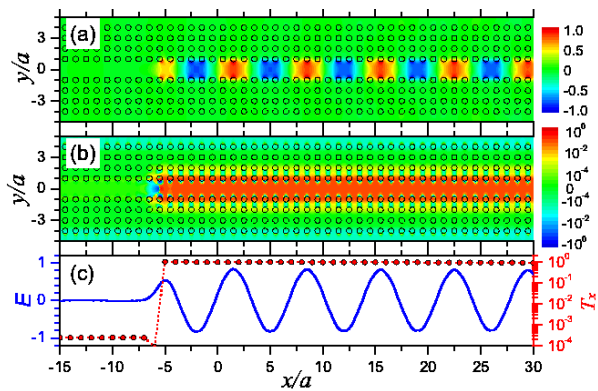}
\caption{\label{fig2} (color online).
The electric field (a) 
and the $x$-component of the Poynting vector (b)
due to a line source of $f=5$ GHz located at $(-5.5a,0)$,
between two MPC slabs with opposite magnetization and $2a$ apart.
Here $a$ is the lattice constant 
of MPC composed of YIG rods indicated by the circles.
Figure (c) displays the electric field $E$ (blue solid line) at $y=0$ and
rightward transmitted power $T_x$ (red dotted line) versus $x$, referred to the left and right
ordinate axes, respectively.
As can be seen, the wave moves only rightward.}
\end{figure}

\begin{figure}
\includegraphics[width=7.9cm]{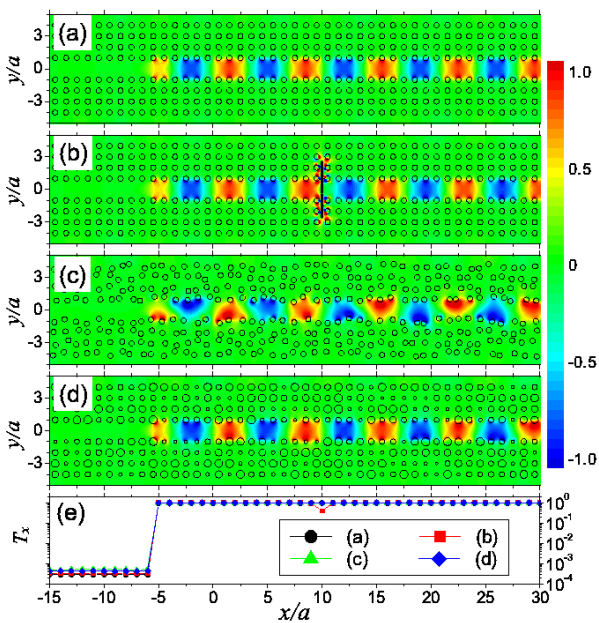}
\caption{\label{fig3} (color online).
(a) The electric field pattern when a line source oscillating at $f=5$ GHz
is placed
between two MPC slabs of opposite magnetization and $2a$ apart, showing
one-way transportation of EM wave.
Introducing a finite linear array of close-packed PEC rods (b), disorder in ferrite rod positions (b),
or disorder in ferrite rod radii (c)
does not obstruct or modify the one-way transportation characteristics,
but, instead, only causes the propagating wave to circumvent the block or
change its wave front, while maintaining complete transmission.
The open circles (of different sizes) denote the YIG rods
while the solid circles in (b) indicate the PEC rods.
Figure (e) shows normalized transmitted power versus $x$ for (a), (b), (c), and (d).}
\end{figure}

An important issue of current interest is the robustness of the
energy flow 
against defects and disorder. One-way propagation
based on the YYR appears immune to scattering from defects and
disorders.
Typical behavior are shown in Fig. 3(b -- d). In Fig. 3(b) we show
the electric field  when a finite linear array of close-packed
perfect electrical conductor (PEC) rods of radii $r_p=\frac18a$ are
inserted into the channel from $y=-2.5a$ to $y=2.5a$, extending over
$2\lambda/3$. The wave circumvents the PEC defect, maintaining
complete power transmission in the channel, with $T_x$ staying
unchanged on the LHS and RHS of the defect, as shown in Fig. 3(e).
The leftward propagating wave caused by the backscattering from the
defect is completely suppressed,
the wave gets around the defect 
and keeps moving rightward. 
By comparing with the waveguide shown in Fig. 3(a), it is found that
the defect changes only the phase of the rightward propagating wave,
partly owing to the delay incurred when it gets around the defect.
In Fig. 3(c) and 3(d), we present the profile of the electric field
for the cases when disorder in the positions and the radii of the YIG rods
is introduced, respectively, in the MPC slab. The change
in position is a random amount from zero to  $25\%$ of the lattice
spacing $a$; the random change in radius has a maximum of $\pm 50$\%
of the unperturbed radius $0.25a$, small enough so that the
disordered rods do not overlap. As can be seen, the disorder, either
in position or in radius, only alters the wave form, but not the
transmission power through the channel, implying the robustness
against disorder, consistent with the fact that the MSP band states
also exhibit finite Chern numbers. For the edge state waveguide
\cite{Haldane, Joan}, the operating frequency lies in the Bragg band
gap, so although their system is immune to scattering from the
defect, it may suffer from disorder in either the position or the
radius of the building blocks in the photonic crystals, as these
effects will destroy the ``Bragg" band gap\cite{msp1}.


Photonic circuits with no back-scattering based on the YYR can be achieved 
even in the deep subwavelength regime.
Figure 4 shows such examples 
where the geometry is scaled down in size with $a=2$ mm and
$r=0.25a$, but the frequency $f=5$ GHz is unchanged.
The other parameters are the same as those in Figs. 3(a). 
A subwavelength guide is shown in 4(a),
with the full lateral
width at half maximum field amplitude $w_h \sim 0.08\lambda$.
In Fig. 4(b), the
configuration is similar to that in Figs. 2 and 3 except that a
cladding slab is added on top of the system.
A line source is located at $(0,-18a)$. The EM wave is 
found to make
a $90^\circ$ turn with nearly $100\%$ power transmission in the
proximity of the MSP resonance frequency, as shown by the blue solid
line in Fig. 4(d). 
If one reverses the orientation of the
magnetization for those rods with coordinates $x<0$ in the cladding
slab, The propagation is seen to split at the bifurcation point,
as shown in Fig. 4(c). The upward transmitted power,
$T_y=\int_{-3a}^{3a} P_y\text{d}x$, is split
with $50\%$ power transmission to the right and the other
$50\%$ to the left by symmetry, as shown by the red dashed line
in Fig. 4(d).
The wave experiences an extremely low reflection at the
corner , as can be seen from the field distribution before bending
and splitting in the vertical channel. Furthermore, as the direction
of the magnetization can be controlled by an external magnetic
field, the function of the system can be switched between the
function of bending and splitting.

\begin{figure}
\includegraphics[width=7.9cm]{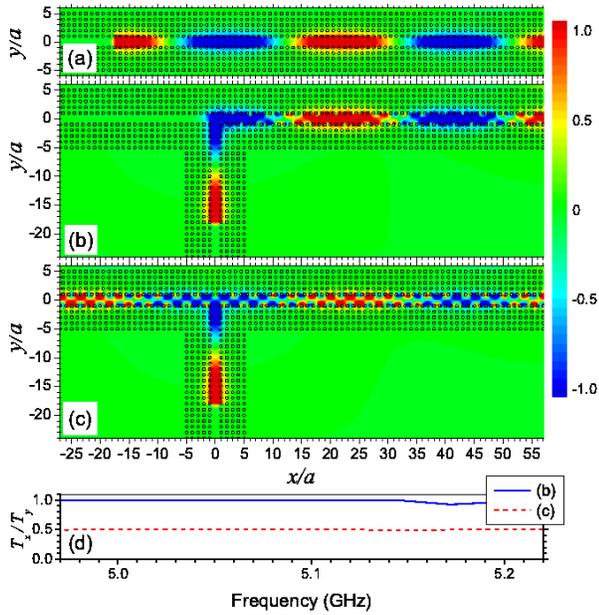}
\caption{\label{fig4} (color online). The electric field patterns
in subwavelength waveguides (a), $90^\circ$ beam bender (b) and splitter (c) 
at $f=5$ GHz. 
The power transmitted horizontally $T_x$ normalized by the 
incoming vertical power, $T_x/T_y$, versus frequency is shown in (d).
The lattice
constant  for the MPC  is $a=2$ mm; the YIG rod radius,
$r=0.25a$. 
The full lateral width at half maximum
field amplitude s $w_h < 0.08\lambda$. }
\end{figure}


In summary, we have demonstrated a new phenomenon, named "Ying-Yang" reflection
when the EM wave is reflected from a MPC. The phenomenon is believed to originate from 
the  TRS breaking nature of the magnetic surface plasmon
band states. 
Possible applications of this phenomenon include a one-way waveguide
that appears to be immune to defect and disorder, sharp wave bending
and beam splitting with extremely low reflection.
These applications are expected to be realizable even in the deep subwavelength scale.  

This work is supported by the China 973 program, NNSFC, PCSIRT, MOE of China (B08011), and
Shanghai Science and Technology Commission.
STC is partly supported by the DOE.

\end{document}